\documentclass[preprint,12pt]{elsarticle}
\journal{Annals of Physics} 
\RequirePackage{graphicx}
\usepackage{color}

\begin{document}

\title{Gravastars with higher dimensional spacetimes}

\author{Shounak Ghosh$^a$, Saibal Ray$^{b}\footnote{$^*$Corresponding author.\\
{\it E-mail addresses:} shounak.rs2015@physics.iiests.ac.in  (SG), saibal@associates.iucaa.in (SR), rahaman@associates.iucaa.in (FR),  bkg@physics.iiests.ac.in (BKG).}$, Farook Rahaman$^c$, B.K. Guha$^a$}

\address{$^a$Department of Physics, Indian Institute of Engineering Science
	and Technology, Shibpur, Howrah, West Bengal, 711103, India\\
$^b$Department of Physics, Government College of Engineering and Ceramic Technology, Kolkata 700010, West Bengal, India\\
$^c$Department of Mathematics, Jadavpur University, Kolkata 700032, West Bengal, India\\
$^a$Department of Physics, Indian Institute of Engineering Science
	and Technology, Shibpur, Howrah, West Bengal, 711103, India}

\date{Received: date / Accepted: date}

\begin{abstract}
We present a new model of gravastar in the higher dimensional
Einsteinian spacetime including Einstein's cosmological constant $\Lambda$.
Following Mazur and Mottola (2001,~2004) we design the star with three specific
regions, as follows: (I) Interior region, (II) Intermediate thin spherical shell and (III) Exterior region.
The pressure within the interior region is equal to the
negative matter density which provides a repulsive force over the shell.
This thin shell is formed by ultra relativistic plasma, where the pressure
is directly proportional to the matter-energy density which does counter balance
the repulsive force from the interior whereas the exterior region is completely vacuum assumed
to be de Sitter spacetime which can be described by the generalized Schwarzschild solution.
With this specification we find out a set of exact non-singular and stable solutions of the gravastar
which seems physically very interesting and reasonable.
\end{abstract}


\maketitle

\section{Introduction}
In general relativity of Einstein there is an inherent feature of singularity
at the end point of gravitationally collapsing system and has been remains an embarrassing
situation to the astrophysical community. To overcome this odd phase of a stellar body
where all the physical laws break down, Mazur and Mottola~\cite{Mazur2001,Mazur2004}
proposed a new model considering the gravitationally vacuum star which was termed
in brevity as Gravastar, that brings up a new arena in the gravitational system.
They generated a new type of solution to this system of gravitational
collapse by extending the idea of Bose-Einstein condensation by constructing
gravastar as a cold, dark and compact object of interior
de Sitter condensate phase surrounded by a thin shell of ultra relativistic
matter whereas the exterior region is completely vacuum, i.e. the Schwarzschild
spacetime is at the outside. The shell is very thin but of finite width in the range
$ R_1 < r < R_2 $, where $R_1=R$ and $R_2=R+\epsilon$ are the interior and exterior
radii of the gravastar, $\epsilon$ represents the thickness of the shell with $\epsilon \ \ll1$.
With this unique specification we can divide the entire system of gravastar
into three specific segments based on the equation of state (EOS) as follows:
(I) Interior: $0 \leq r < R_1 $,~with EOS~$ p = -\rho $, (II) Shell: $ R_1 \leq r \leq R_2 $,
~with EOS~$ p = +\rho $, and (III) Exterior: $ R_2 < r $,~with EOS~$ p = \rho =0$.

The abovementioned model of gravastar has been studied by researchers
which opened up a new challenges in the gravitational research to obtain
a singularity free solution of the Einstein field equations.  Therefore, it is
supposed to be an alternative solution of black hole and has been studied by
several authors in different context of astrophysical systems~\cite{Visser2004,
Cattoen2005,Carter2005,Bilic2006,Lobo2006,DeBenedictis2006,
Lobo2007,Horvat2007,Cecilia2007,Rocha2008,Horvat2008,Nandi2009,
Turimov2009,Usmani2011,Rahaman2012a,Rahaman2012b,Rahaman2012c,Ghosh2017}.

The negative matter density in the interior region creates
a repulsive pressure acting radially outward from the centre of the
gravastar $(i.e.~r=0)$ over the shell whereas the shell of positive
matter density provides the necessary gravitational pull to
balance this repulsive force within the interior. It is assumed that
the dark energy (or the vacuum energy) is responsible for this repulsive
pressure from the interior. In a general consideration, the EOS $ p = -\rho $
is suggesting that the repulsive pressure is an agent,
responsible for accelerating  phase of the present universe and is known as the $\Lambda$-dark
energy~\cite{Riess1998,Perlmutter1999,Ray2007b,Usmani2008,Frieman2008}. In literature this
EOS is termed as a `false vacuum', `degenerate vacuum', or `$\rho$-vacuum'~\cite{Davies1984,Blome1984,Hogan1984,Kaiser1984}.
 Therefore, in this context one can note that gravastar may have some connection to the dark
star~\cite{Lobo2008,Chan2009a,Chan2009b,Usmani2011}.

The EOS for the shell $p=\rho$ represents essentially a stiff fluid model as conceived by Zel'dovich~\cite{Zeldovich1972}
in connection to the cold baryonic universe. The idea has been considered by several scientists for various situations in cosmology~\cite{Carr1975,Madsen1992} as well as astrophysics~\cite{Buchert2001,Braje2002,Linares2004}.

The so-called cosmological constant $\Lambda$ was introduced by Einstein in his field equations to obtain a static and non-expanding solutions of the universe. In his model this constant with the right sign could produce a repulsive pressure to exactly counter balance the gravitational attraction and thus could provide a statical model of the universe.

After the experimental verification of expanding universe by Edwin Hubble between 1922 to 1924~\cite{Hubble1929} and the success of FLRW cosmology made Einstein realize that the universe has been expanding with an acceleration. That is why later on Einstein discarded the cosmological constant from his field equation. However, though it is abandoned by Einstein but for the physical requirement to describe
one-loop quantum vacuum fluctuations, the Casimir effect~\cite{Casimir1948}, cosmological constant had to appear once again in the theory with a form as $T_{ij}=\Lambda g_{ij}/{8 \pi G}$, where $T_{ij}$ and $g_{ij}$ are the stress energy tensor and the metric tensor
respectively and $G$ is the usual Newtonian constant.

Recent observations conducted by WMAP suggests that $73\%$ of the total mass-energy
of the universe is dark energy~\cite{Ruderman1972,Perlmutter1998}. It is believed that
this dark energy plays an important role for the evolution of the universe and
in order to describe the dark energy scientists have recall the erstwhile cosmological constant.
Therefore, in the modern cosmology this cosmological constant is treated as a strong
candidate for the dark energy which is responsible for the accelerating phase
of the present universe.

It is observed that in superstring theory spacetime is considered to be of dimensions higher than four. To be consistent with the
physically acceptable models the $4$-dimensional present spacetime fabric is assumed to be self-compactified form of manifold with
multidimensions or extra spatial dimensions~\cite{Schwarz1985,Weinberg1986,Duff1995,Polchinski1998,Hellerman2007,Aharony2007}.
Following this aspect very recently Bhar~\cite{Bhar2014} has proposed a charged gravastar in higher dimension by admitting conformal motion and later on worked out a problem by Ghosh et al.~\cite{Ghosh2017} without admitting the conformal motion in the framework of Mazur and Mottola model. Some authors~\cite{Pradhan2007,Emparan2008,Khadekar2014} have studied higher dimensional works admitting one parameter Group of Conformal motion. Usmani et al.~\cite{Usmani2011} have also found solution of neutral gravastar in higher dimension without admitting the conformal motion. These works provide an alternative solution to the static black holes. The present study on gravastar basically is an extension of the work of Usmani et al.~\cite{Usmani2011} and its generalization to the higher dimensional spacetime in presence of the cosmological constant $(\Lambda)$.

Therefore, the main motivation of this work is to study the effects of the cosmological constant for construction of gravastars and also to study the higher dimensional effects, if any. The present investigations are based on the plan as follows: The background of the model has been implemented through the Einstein geometry in Sec. 2, whereas the solution of interior spacetime, the thin shell and exterior spacetime of the gravastar have been discussed in Sec. 3. We have shown the matching condition in Sec. 4 and discussed the junction conditions for the different regions of the gravastar in Sec. 5. Some physical features of the model, viz. the proper length, Energy, Entropy are explored in the Sec. 6 which followed by the discussions and concluding remarks in Sec. 7.

\section{The Einsteinian relativistic spacetime geometry}
The Einstein-Hilbert action coupled to matter is given by
\begin{equation}
I = \int d^D x \sqrt{-g } \left( \frac{R_D}{16 \pi G_D} + L_{m}\right),\label{eq1}
\end{equation}
where the curvature scalar in $D$-dimensional spacetime is represented by $R_D$,
with $G_D$ is the $D$-dimensional Newtonian constant and $L_{m}$ denotes
the Lagrangian for the matter distribution. We obtain the following
Einstein equation by varying the above action with respect to the
metric
\begin{equation}
G^D_{ij}   = - 8 \pi G_D T_{ij},\label{eq2}
\end{equation}
where $G^D_{ij}$ denotes the Einstein's tensor in $D$-dimensional
spacetime.

The interior of the star is assumed to be perfect fluid type and
can be given by
\begin{equation}
T_{ij} = (\rho + p ) u_i u_j + p  g_{ij}, \label{eq3}
\end{equation}
where $\rho$ represents the energy density, $p$ is  the isotropic
pressure, and  $u^{i}$ is the $D$-velocity of the fluid.

Here in the present study it is assumed that the gravastars in higher dimensions
have the $D$-dimensional spacetime with the structure $R^1 X S^1 X S^d\ (d = D - 2)$,
where the range of the radial coordinate is $S^1$ and the time axis is represented by $R^1$.
For this purpose, we consider a static spherically symmetric metric in $D = d + 2$ dimension as
\begin{equation}
ds^2 = -e^{\nu}dt^2 + e^{\lambda}dr^2+r^2 d\Omega_d ^2,\label{eq4}
\end{equation}
where $ d\Omega_d ^2$ is the linear element of a
$d$-dimensional unit sphere, parameterized by the angles $\phi_1,
\phi_2,......,\phi_d$ as follows:\\

$d\Omega_d ^2= d\phi_d^2 + \sin_2 \phi_d [d\phi_{d-1}^2 + \sin_2
\phi_{d-1}\{d\phi_{d-2}^2 + .........+ \sin_2 \phi_3(d\phi_2^2 +
\sin_2 \phi_2  d\phi_1^2).......\}] $.

Now the Einstein field equations for the metric (\ref{eq4}), together
with the energy-momentum tensor in presence of the non-zero cosmological constant $\Lambda$, yield
\begin{eqnarray}
-e^{-\lambda} \left[\frac{d(d-1)}{2r^2} - \frac{d \lambda'}{2r}
\right] + \frac{d(d-1)}{2r^2} = 8\pi G_D~ \rho + \Lambda,~~~~~~~~~~~~~~~~~~~~~~~~~~~~~~~~~ \label{eq5}\\
e^{-\lambda} \left[\frac{d(d-1)}{2r^2} + \frac{d \nu'}{2r} \right]
- \frac{d(d-1)}{2r^2} = 8\pi G_D ~p -\Lambda, ~~~~~~~~~~~~~~~~~~~~~~~~~~~~~~~~~~\label{eq6}\\
\frac{e^{-\lambda}}{2} \left[ \nu'' -\frac{\lambda'\nu'}{2}
+\frac{ {\nu'}^2}{2} -\frac{(d-1)(\lambda'-\nu')}{r} +
\frac{(d-1)(d-2) }{r^2} \right]- \frac{(d-1)(d-2)
}{2r^2} \nonumber \\  = 8\pi G_D~p -\Lambda,\label{eq7}~~~~~~~~~~~~
\end{eqnarray}
where `$\prime$' denotes differentiation with respect to the radial parameter $r$. Following geometrical unit system we
have assumed $c = 1$ throughout the paper.

The general relativistic conservation of energy-momentum, ${T^{ij}}_;j=0$,
can be expressed in its general form with $D$-dimension as
\begin{equation}
\frac{1}{2} \left(\rho + p\right)\nu' + p' =0 . \label{eq8}
\end{equation}

In the next Sec. 3 we shall formulate special explicit forms of the
energy conservation equations for all the three regions,
viz. interior, intermediate thin shell and exterior spacetimes.

\section{Modelling a gravastar under general relativity}

\subsection{Interior spacetime}
The interior region of the gravastar is so designed that the negative pressure which is
acting radially outward from the central part of the star could balance
the inward gravitational pulling from the shell of the spherical system. To fulfil this criterion,
we choose the EOS for the interior region in the following form~\cite{Mazur2001},
\begin{equation}
p =- \rho. \label{eq9}
\end{equation}

Using  Eq. (\ref{eq8}) and the above EOS  (\ref{eq9}), we obtain
\begin{equation}
p =-\rho =- \rho_c, \label{eq10}
\end{equation}
where $\rho_c$ is the constant density of the interior region.

Using Eq. (\ref{eq10}) in the field equation (\ref{eq5}), one obtains
the solution of $\lambda$ as
\begin{equation}
e^{-\lambda} = 1-\frac{16\pi G_D\rho_c}{d(d+1)}r^2 -\frac{2\Lambda r^2}{d(d+1)}+ C_1r^{1-d},\label{eq11}
\end{equation}
where  $C_1$ is an integration constant. We observe from the fourth factor of the above expression that there involves a constraint $d \neq 1$ which means we must consider $d\geq2$. Hence the solution to be regular at $r=0$, one can demand for $C_1=0$. Thus essentially we get
\begin{equation}
e^{-\lambda} = 1-\frac{16\pi G_D\rho_c}{d(d+1)}r^2-\frac{2\Lambda r^2}{d(d+1)}.\label{eq12}
\end{equation}

The variation of $e^{-\lambda}$ with respect to the radial coordinate $r$ is plotted in Fig. 1.

\begin{figure*}[!htp]\centering
\includegraphics[width=0.5\textwidth]{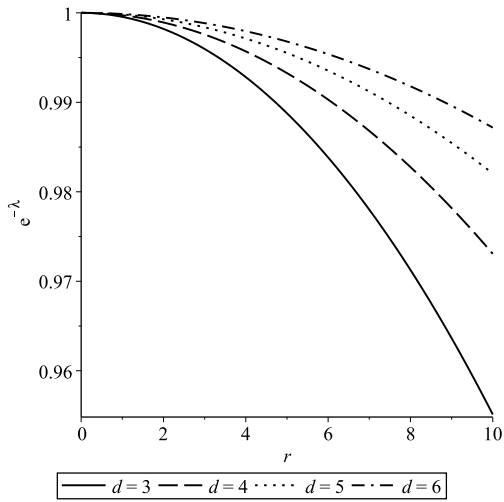}
\caption{Variation of the $e^{-\lambda}$ with the radial coordinate $r$ (km) for different
dimensions in the interior region where the specific legends used are shown in the
respective plots}
\end{figure*}

Now by employing Eq. (\ref{eq9}) in Eqs. (\ref{eq5}) and (\ref{eq6}), we obtain
\begin{equation}
  \ln k =\lambda + \nu \Rightarrow e^{\nu}=ke^{-\lambda}, \\ \label{eq13}
\end{equation}
where $k$ is a constant of integration.

Thus we have the following interior solutions for the metric
potentials $\lambda$ and $\nu$ as follows
\begin{equation}
 ke^{-\lambda} =e^\nu =k\left[1-\frac{16\pi G_D\rho_c}{d(d+1)}r^2- \frac{2\Lambda }{d(d+1)}r^2\right],\label{eq14}
\end{equation}

From Eq. (\ref{eq10}) it is observed that the matter density remains constant over the entire interior spacetime.
Thus we can calculate the active gravitational mass $M(r)$ in higher dimensions as
\begin{equation}
M(r) = \int_0^{{R_1=R}}~ \left[  \frac{2 \pi^{\frac{d+1}{2}}
}{\Gamma \left(\frac{d+1}{2}\right)}\right]r^d \rho_c dr
\\ = \left[ \frac{2 \pi^{\frac{d+1}{2}} \rho_c}{(d+1)\Gamma
\left(\frac{d+1}{2}\right)}\right]
R^{d+1},\label{eq15}
\end{equation}
where $R$ is the internal radius of the gravastar. So the usual gravitational mass for a $d$-dimensional sphere
can be expressed by Eq. (\ref{eq15}). This explicitly shows that the active gravitational mass $M(r)$ is directly
dependent on the following two factors - the radius $R$ as well as the matter density $\rho$.

\subsection{Intermediate thin shell}
Here we assume that the thin shell contains ultra-relativistic fluid of soft quanta and obeys the following EOS
\begin{equation}
p =\rho. \label{eq17}
\end{equation}

By inspection we note that it is very difficult to obtain a general solution of the Einstein field equations within the shell of non-vacuum region. Therefore, we try to find an analytic solution within the thin shell limit,  $0<e^{-\lambda}\equiv h <<1$.  To do so we set $h$ to be zero to the leading order. Under this approximation, the field Eqs. (\ref{eq5}) - (\ref{eq7}) along with the above EOS, can be reformatted in the following form
\begin{eqnarray}
&&~~~~~~~~~~~~~~~~~~~~~~~~~~-\frac{h'}{2 r}  + \frac{(d-1)}{2r^2}=\frac{8\pi G_D\rho}{d}+\frac{\Lambda}{d},  \label{eq18} \\
&&~~~~~~~~~~~~~~~~~~~~~~~~~~~~~~ -\frac{(d-1)}{2r^2}= \frac{8\pi G_Dp}{d}-\frac{\Lambda}{d}, \label{eq19} \\
&&\frac{h'\nu'}{4}+\frac{(d-1)h'}{2r} -\frac{(d-1)(d-2)}{2r^2}=8\pi G_D p-\Lambda. \label{eq20}
\end{eqnarray}

Now using Eqs. (\ref{eq18}) and (\ref{eq19}), we find out an expression for $h$ as
\begin{equation}
h=e^{-\lambda} = k_{1}+ 2 (d-1) \ln r - \frac{2 \Lambda r^2}{d},\label{eq21}
\end{equation}
where $k_{1}$ is an integration constant. It is to note that the range of $r$ lies within the thickness of the shell, i.e. $R_1=R$ and $R_2=R+\epsilon$, where $\epsilon$ the thickness of the shell ($\epsilon <<1$).

The other metric potential ($\nu$) can be found as
\begin{equation}
~~~~~~~~~~~~~~~~~~~~e^{-\nu}=k_2 \left[\frac{r^{2d}}{2\Lambda r^2 -d^2+d}\right],\label{eq22}
\end{equation}
where $k_2$ is the integration constant.

Again using Eqs. (\ref{eq8}) and (\ref{eq22}) we get the pressure and matter density
within the shell of the gravastar as
\begin{equation}
~~~~~~~~~~~~~~~~p=\rho=\rho_0e^{-\nu}=\rho_0 k_2 \left[\frac{r^{2d}}{2\Lambda r^2 -d^2+d}\right].\label{eq23}
\end{equation}

The variation of $p=\rho$ with respect to the radial coordinate $r$ is plotted in Fig. 2.

\begin{figure*}[thbp]
\centering
\includegraphics[width=0.5\textwidth]{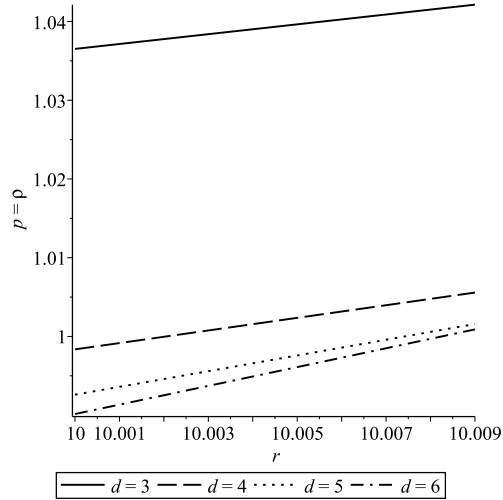}\\
\caption{Variation of the pressure $p=\rho$ (km$^{-2}$) of the ultra
relativistic matter in the shell with the radial coordinate $r$ (km) for different
dimensions}
\end{figure*}

In the intermediate thin shell, therefore, by virtue of Eq. (\ref{eq23}) the energy conservation equation (\ref{eq8}) takes the special explicit form in the pressure gradient as follows:
\begin{equation}
p'=2\rho_0 k_2 \left[\frac{r^{2d}}{2\Lambda r^2 -d^2+d}\right]\left[\frac{2\Lambda r}{2\Lambda r^2-d^2+d}-\frac{d}{r}\right].\label{eq24}
\end{equation}

\subsection{Exterior spacetime}
The EOS for the exterior region is defined as $p=\rho=0$. In higher dimensions
one can expect that the exterior solution is nothing but a generalization of the Schwarzschild solution.
Now, following the work of Tangherlini~\cite{Tangherlini1963} this can be obtained
as
\begin{equation}
ds ^2 = - \left(1 - \frac{\mu}{ r^{d-1}}-\frac{ 2\Lambda r^2}{(d+1)d}\right)
dt^2 + \left(1 - \frac{\mu}{ r^{d-1}}-\frac{ 2\Lambda r^2}{(d+1)d}\right)^{-1} dr^2 +r^2 d \Omega_d^2,\label{eq25}
\end{equation}
where $\mu$ is a constant and is given by $\mu={16\pi G_D M}/{\Omega_d}$ in higher
dimension, with $M$ as the mass of the gravastar and ${\Omega}_d$ as the area
of a unit $d$-sphere which is defined by ${\Omega}_d ={2 \pi^{(\frac{d+1}{2})}}/{\Gamma(\frac{d+1}{2})}$.

However, Eq. (\ref{eq25}) appears to have a cosmological constant $p = -\rho$ and hence due to the EOS $p=\rho=0$, the above metric eventually reduces to
\begin{equation}
ds ^2 = - \left(1 - \frac{\mu}{ r^{d-1}}\right)
dt^2 + \left(1 - \frac{\mu}{ r^{d-1}}\right)^{-1} dr^2 +r^2 d \Omega_d^2.\label{eq25a}
\end{equation}

\section{Matching condition for finding out expressions of constants}
For the construction of the gravastar the metric potential $g_{rr}$ must be continuous at the interface between the core and the shell at $r=R_1$ (interior radius) and also at junction of the shell and the exterior region at $r=R_2$ (exterior radius). Using this matching condition one can obtain the values of the integration constants $k_1$ and $k_2$ as shown below.

To find the value of $k_1$ we gave matched the metric potential at $r=R_1$.
Using Eq. (\ref{eq12}) and (\ref{eq21})we have the value of $k_1$ as
\begin{equation}
k_1=1-\frac{16\pi G_D\rho_c}{d(d+1)}R_1^2+\frac{2\Lambda R_1^2}{d+1}-2(d-1)\ln R_1.\label{eq27}
\end{equation}

Again from Eqs. (\ref{eq22}) and (\ref{eq25}) at $r=R_2$ we get the value of $k_2$ as
\begin{equation}
k_2=\frac{2\Lambda R_2^2 -d^2+d}{{R_2^{2d}}\left(1 - \frac{\mu}{ R_2^{d-1}}-\frac{ 2\Lambda R_2^2}{(d+1)d}\right)},\label{eq28}
\end{equation}

We choose the numerical values of $\mu=3.75M_\odot$, $R_1$=10~km and $R_2$=10.009~km.

\section{Junction condition}
There are three regions in the gravastar configuration, viz.
the interior region, thin shell and exterior region. The shell joins the
interior and exterior regions at the junction interface. The metric
coefficients are continuous at the shell, however we do not have any confirmation
of the continuity of their derivatives.

Following the condition of Darmois-Israel~\cite{Darmois1927,Israel1966} now we
provide the intrinsic surface stresses at the junction interface in the form
\begin{equation}
S_{ij}=-\frac{1}{8\pi}(\kappa_{ij}-\delta_{ij}\kappa_{kk}),\label{eq29}
\end{equation}
where $k_{ij}=K^+_{ij}-K^-_{ij}$, that shows the discontinuity of the
extrinsic curvatures or second fundamental forms. Here the signatures $-$ and $+$
describes the interior and exterior boundaries respectively of the gravastar.

Now this extrinsic curvature connect the two sides of the thin shell as
\begin{equation}
K_{ij}^{\pm}=-n_{\nu}^{\pm}\left[\frac{\partial^{2}X_{\nu}}{\partial
\xi^{1}\partial\xi^{j}}+\Gamma_{\alpha\beta}^{\nu}\frac{\partial
X^{\alpha}}{\partial \xi^{i}}\frac{\partial X^{\beta}}{\partial
\xi^{j}} \right]|_S, \label{eq30}
\end{equation}
where $n_{\nu}^{\pm}$ is the unit normals to the surface $S$ can be defined as
\begin{equation}
n_{\nu}^{\pm}=\pm\left|g^{\alpha\beta}\frac{\partial f}{\partial
X^{\alpha}}\frac{\partial f}{\partial X^{\beta}}
\right|^{-\frac{1}{2}}\frac{\partial f}{\partial X^{\nu}}, \label{eq31}
\end{equation}
with $n^{\nu}n_{\nu}=1$.

Following the methodology of Lanczos~\cite{Lanczos1924} one can obtain the surface energy-momentum
tensor on the thin shell as $S_{ij}=diag[{-\Sigma, p_{\theta_1}, p_{\theta_2}...,~p_{\theta_d}}$],
where $\Sigma$ is the surface energy density and $ p_{\theta_{1}}=p_{\theta_{2}}=...=p_{\theta_{d}}=p_t$
are the surface pressures which respectively can be determined by
\begin{eqnarray}
\Sigma&=&-\frac{d}{8\pi R}\sqrt{f} \label{eq32}\\
p_t&=-&\frac{d-1}{d}\Sigma +\frac{f^{'}}{16\pi\sqrt{f}}. \label{eq33}
\end{eqnarray}

Using the above equations one can obtain
\begin{equation}
\Sigma=-\frac{d}{8\pi R}\left[\sqrt{1 - \frac{\mu}{ R^{d-1}}-\frac{2R^2 \Lambda}{(d+1)d}}-
\sqrt{ 1-\frac{16\pi G_D\rho_c}{d(d+1)}R^2-\frac{2\Lambda R^2}{d(d+1)}}\right] \label{eq34}
\end{equation}
and
\begin{eqnarray}
p_t=\frac{1}{8\pi R}\left[\frac{(d-1)-\frac{(d-1)\mu}{2R^{d-1}}-\frac{2\Lambda R^2}{d+1}}{\sqrt{1 - \frac{\mu}{ R^{d-1}}-\frac{2R^2 \Lambda}{(d+1)d}}}-
\frac{(d-1)-\frac{16\pi G_D\rho_c}{d+1}R^2-\frac{2\Lambda R^2}{d+1}}{\sqrt{ 1-\frac{16\pi G_D\rho_c}{d(d+1)}R^2-\frac{2\Lambda R^2}{d(d+1)}}}\right]. \label{eq35}
\end{eqnarray}

Now, it is easy to find out the mass $m_s$  of the thin shell as
\begin{equation}
m_s=\frac{2\pi^{\frac{d+1}{2}}}{\Gamma(\frac{d+1}{2})}R^d \Sigma= \Omega_d' R^d \left(\sqrt{ 1-\frac{16\pi G_D\rho_c R^2}{d(d+1)}-\frac{2\Lambda R^2}{d(d+1)}}-\sqrt{1 - \frac{\mu}{ R^{d-1}}-\frac{2R^2 \Lambda}{d(d+1)}} \right), \label{eq36}
\end{equation}
where $\Omega_d'=\frac{2\pi^{\frac{d+1}{2}}}{\Gamma(\frac{d+1}{2})}\frac{d}{8\pi R}$.

Using Eq. (\ref{eq36}) we can determine the total mass of the gravastar in terms of the mass of the thin shell in the form
\begin{equation}
\mu=\frac{m_s }{\Omega_d' R}
\left[2\sqrt{ 1-\frac{16\pi G_D\rho_c}{d(d+1)}R^2-\frac{2\Lambda R^2}{d(d+1)}}-\frac{m_s}{\Omega_d' R^d}\right] + \frac{16 \pi G_D \rho_c }{d(d+1)}R^{d+1}. \label{eq37}
\end{equation}

\section{Physical features of the models}

\subsection{Energy content}
As we consider the thickness of the intermediate shell is very
small $(0 < \epsilon \ll 1)$, so the phase boundary can be described by
the interfaces at $r=R$ and $r=R+\epsilon$ joining the region I and
region III respectively, where $R$ describes the phase boundary of the region I.

Using the Eq. (\ref{eq23}) we calculate the energy within the shell as
\begin{eqnarray}
E  =   \int _{R}^{R+\epsilon}\left[\frac{2 \pi^{\frac{d+1}{2}} }
{\Gamma \left(\frac{d+1}{2}\right)}\right]r^d \rho dr
&=& \left[ \frac{2 \pi^{\frac{d+1}{2}} }{\Gamma\left(\frac{d+1}{2}\right)}\right]\rho_0 k_2 \int _{R}^{R+\epsilon}  \left(\frac{r^{3d}}{2\Lambda r^2 -d^2+d}\right). \nonumber\\ \label{eq38}
 \end{eqnarray}

To solve the above integration of Eq. (\ref{eq38}) over the limit $R$ to $R+\epsilon$, let us form
a special differential equation in terms of the parameter $F(r)$ which can be represented as
 \begin{equation}
 \frac{dF(r)}{dr}=\frac{r^{3d}}{2\Lambda r^2 -d^2+d}. \label{eq39}
 \end{equation}

Now using the above consideration we can solve the the integration as follows
\begin{eqnarray}
\int _{R}^{R+\epsilon}\frac{dF(r)}{dr}dr &=&[F(r)]_R^{R+\epsilon} = F(R+\epsilon)-F(R) \approx \epsilon \left(\frac{dF}{dr}\right)_{r=R}. \label{eq40}
 \end{eqnarray}

As $\epsilon\ll 1$, so we consider up to the first order term of the Taylor series expansion for the expression in Eq. (\ref{eq40}).

Therefore, combining Eqs. (\ref{eq38}) and (\ref{eq40}) we get
\begin{eqnarray}
E \approx \left[\frac{2 \pi^{\frac{d+1}{2}} }{\Gamma\left(\frac{d+1}{2}\right)}\right]  \left(\frac{\epsilon \rho_0 k_2 R^{3d}}{2\Lambda R^2 -d^2+d}\right),\label{eq41}
\end{eqnarray}
where in the square bracket of Eqs. (\ref{eq38}) and (\ref{eq41}) we have used the factor ${\Omega}_d$ as the surface area
of a unit $d$-sphere which has already been defined by ${\Omega}_d =2 \pi^{\frac{d+1}{2}}/\Gamma(\frac{d+1}{2})$.

From Eq. (\ref{eq41}) we observe that the energy within the shell is directly proportional to the thickness of the
shell $(\epsilon)$.

\begin{figure*}[thbp]
\centering
\includegraphics[width=0.5\textwidth]{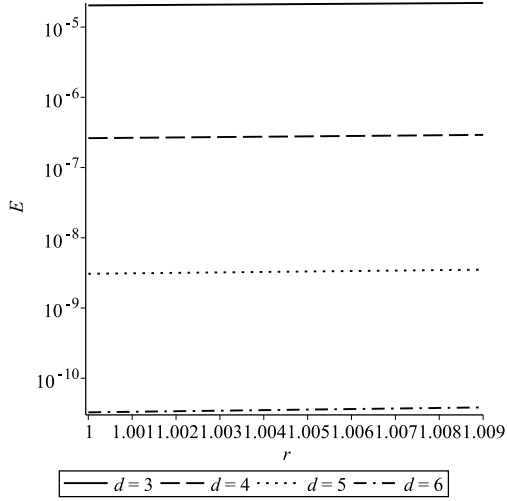}
\caption{Variation of the energy E (km) of the shell with $r$ (km) for different dimensions }
\end{figure*}

The variation of $E$ with $r$ for different dimensions is shown in Fig. 3.
One can note that the energy is increasing from the interior boundary to
the exterior boundary. This is clearly indicating that the shell is getting
harder from the interior to exterior boundary, which suggests that the
exterior boundary is more dense than the interior boundary as obtained
in Fig. 2. The plot also indicates that energy is decreasing as
the dimension increases.

\subsection{Entropy}
The entropy within the shell in higher dimensions can be obtained using the following equation
\begin{equation}
 S=\int^{R+\epsilon}_R \frac{2\pi^{\frac{d+1}{2}}}{\Gamma(\frac{d+1}{2})} r^d s(r)\sqrt{e^{\lambda}} dr,\label{eq42}
\end{equation}
where $s(r)$ is the entropy density, following the prescription Mazur and Mottola \cite{Mazur2001} we can write it as follows
\begin{equation}
 s(r)=\frac{\xi^2 k_B^2 T(r)}{4\pi\hbar^2}=\xi\frac{k_B}{\hbar}\sqrt{\frac{p}{2\pi}} \label{eq43}
\end{equation}
where $\xi$ is a dimensionless constant and $T$ is the local temperature which depends on the radial coordinate $r$.

By using the above equation Eq. (\ref{eq43}) in Eq. (\ref{eq42}), we have
\begin{equation}
S =  \frac{2\pi^{\frac{d+1}{2}}}{\Gamma(\frac{d+1}{2})}\frac{\xi k_B}{\hbar }\sqrt{\frac{\rho_0 k_2}{2 \pi}}\int_R^{R+\epsilon} \frac{{r^{2d}}}{\sqrt{(2\Lambda r^2 - d^2 + d)[k_1+2(d-1)\ln r-\frac{2\Lambda r^2}{d}]}}. \label{eq44}
\end{equation}

The above integration can be solved for small thickness limit $(\epsilon \ll 1)$
by using the Taylor series expansion and we obtain
\begin{equation}
S \approx \frac{2\pi^{\frac{d+1}{2}}}{\Gamma(\frac{d+1}{2})}\frac{\xi k_B}{\hbar }\sqrt{\frac{\rho_0 k_2}{2 \pi}}
\frac{{\epsilon R^{2d}}}{\sqrt{(2\Lambda R^2 - d^2 + d)[k_1+2(d-1)\ln R-\frac{2\Lambda R^2}{d}]}}. \label{eq45}
\end{equation}

\begin{figure*}[thbp]
\centering
\includegraphics[width=0.5\textwidth]{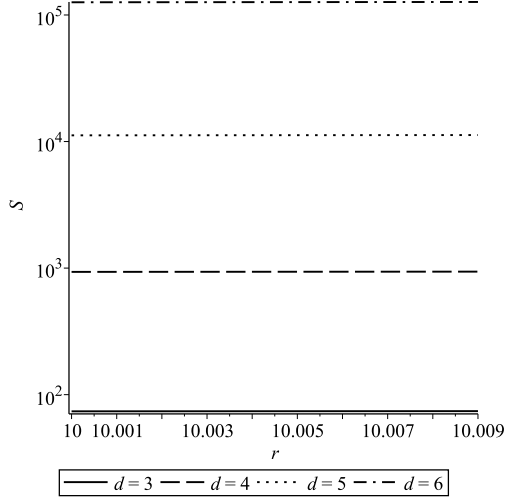}
\caption{Variation of the entropy of the shell with $r$ (km) for different dimensions }
\end{figure*}

It is observed from Eq. (\ref{eq45}) that the entropy within the shell is directly
proportional with the thickness of the shell. The variation of the entropy $(S)$
with $r$ for different dimensions shown in Fig. 4 and it shows almost similar
in nature as the variation of energy obtained in Fig. 3.

\subsection{Proper length}
Now, the proper length between two interfaces of the shell can be
calculated by using the following equation
\begin{eqnarray}
\ell = \int _{R}^{R+\epsilon} \sqrt{e^{\lambda} } dr
&=& \int_{R}^{R+\epsilon} \frac{dr}{{\sqrt{k_{1}+ 2 (d-1) \ln r - \frac{2\Lambda r^2}{d}}}}\nonumber\\
 &\approx& \frac{\epsilon}{\sqrt{k_{1}+ 2 (d-1) \ln R - \frac{2\Lambda R^2}{d}}}.\nonumber\\ \label{eq46}
\end{eqnarray}

Variation of the proper length has been shown in Fig. 5.

\begin{figure*} [thbp]
\centering
\includegraphics[width=0.5\textwidth]{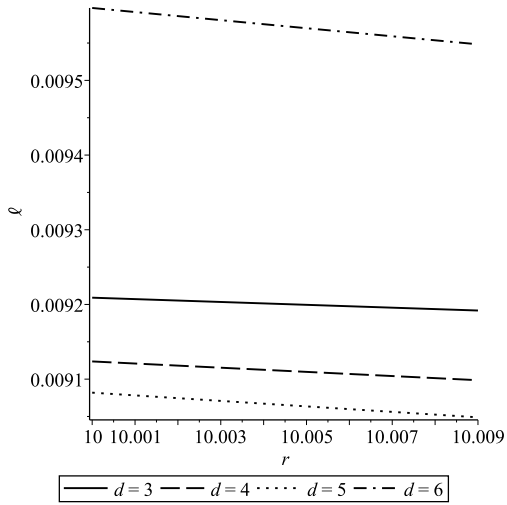}
\caption{Variation of the proper length of the shell with the radial coordinate $r$ (km) for different
dimensions}
\end{figure*}

\subsection{Equation of State}
 Let us assume that $p_{\theta_{1}}=p_{\theta_{2}}=p_{\theta_{3}}=...=p_t=-T$,
where $T$ is the surface tension. Therefore, Eqs. (\ref{eq34}) and (\ref{eq35}) yield
\begin{equation}
T=\omega(R)\Sigma. \label{eq47}
\end{equation}

Thus, the EOS parameter can be found as
\begin{equation}
\omega(R)=-\frac{\left[\frac{\frac{d-1}{d}-\frac{(d-1)\mu}{2dR^{d-1}}-\frac{2\Lambda R^2}{d(d+1)}}{\sqrt{1 - \frac{\mu}{ R^{d-1}}-\frac{2R^2 \Lambda}{(d+1)d}}}-\frac{\frac{d-1}{d}-\frac{16\pi G_D\rho_c}{d(d+1)}R^2-\frac{2\Lambda R^2}{d(d+1)}}{\sqrt{ 1-\frac{16\pi G_D\rho_c}{d(d+1)}R^2-\frac{2\Lambda R^2}{d(d+1)}}}\right].}{\left[\sqrt{1 - \frac{\mu}{ R^{d-1}}-\frac{2R^2 \Lambda}{(d+1)d}}-\sqrt{ 1-\frac{16\pi G_D\rho_c}{d(d+1)}R^2-\frac{2\Lambda R^2}{d(d+1)}}\right]}.\label{eq48}
\end{equation}

\subsection{Stability}
To check the stability of the gravastar in higher dimensions we are following the prescription suggested by Mazur and Mottola~\cite{Mazur2001,Mazur2004}. For doing this we have maximized the entropy functional with respect to all variations of the mass function $\mu(r)$ and investigate the sign of second variation of the entropy functional. As the end points of the boundary are fixed at $R_1$ and $R_2$ respectively, so the first variation of $S$ must vanish at $R_1$ and $R_2$, i.e., $\delta S=0$ by Einstein's field equations of Eq. (\ref{eq5}) to Eq. (\ref{eq7} for a static, spherically symmetric star.

Now, the higher dimensional generalization of the entropy functional can be obtained as
\begin{equation}
S =  \sqrt{\frac{2\pi^{\frac{d+1}{2}}}{\Gamma(\frac{d+1}{2})}}\frac{\xi k_B}{2 \hbar \sqrt{2\pi}}\int^{R_2}_{R_1} r^{\frac{d}{2}}dr \left(\frac{d\mu}{dr}\right)^{\frac{1}{2}} h^{-1/2}, \label{eq49}
\end{equation}
where $h=1-\frac{\mu}{r^{d-1}}-\frac{2\Lambda r^2}{d(d+1)}$.

Now from Eqs. (\ref{eq21}), (\ref{eq25}) and (\ref{eq27}) we get the form of $\mu(r)$ as
\begin{equation}
 \mu(r)=\frac{16\pi G_D \rho_c r^{d-1}R_1^2}{d(d+1)}+\frac{2\Lambda r^{d+1}}{d+1} -\frac{2\Lambda R_1^2 r^{d-1}}{d+1}+2(d-1)r^{d-1}\ln{\frac{R_1}{r}}. \label{eq50}
\end{equation}

Hence we can have second variation of the entropy function as
\begin{eqnarray}
\delta^2 S &=&  \sqrt{\frac{2\pi^{\frac{d+1}{2}}}{\Gamma(\frac{d+1}{2})}}\frac{\xi k_B}{8 \hbar \sqrt{2\pi}}\int^{R_2}_{R_1} r^{\frac{d}{2}}dr \left(\frac{d\mu}{dr}\right)^{\frac{-3}{2}} h^{-1/2}\left[-\left(\frac{d(\delta \mu)}{dr}\right)^2+\nonumber \right.\\
 && \left.\frac{(\delta \mu)^2}{h^2r^{2(d-1)}}\left(\frac{d\mu}{dr}\right)\left(1+\frac{d\mu}{dr}\right)\right]. \label{eq50a}
\end{eqnarray}

As a next step we consider the following linear combination $\delta \mu = \chi \psi$, where $\psi$ does vanish at the endpoints. Now insert this into the second variation (Eq. \ref{eq50a}), after integrating by parts with $\delta \mu =0$ at the extreme points of the shell, we are left with
\begin{equation}
\delta^2 S = - \sqrt{\frac{2\pi^{\frac{d+1}{2}}}{\Gamma(\frac{d+1}{2})}}\frac{\xi k_B}{8 \hbar \sqrt{2\pi}}\int^{R_2}_{R_1} r^{\frac{d}{2}}dr \left(\frac{d\mu}{dr}\right)^{\frac{-3}{2}} h^{-1/2}\chi^2\left[\frac{d\psi}{dr}\right]^2 < 0. \label{eq50b}
\end{equation}

It clearly indicates that the entropy function with higher dimensions has the maximum value for all radial variations which are vanished at extreme points of the boundary of the shell. Therefore, following Mazur and Mottola~\cite{Mazur2004} we can conclude that ``perturbations of the fluid in region II that are not radially symmetric decrease the entropy..., which demonstrates that the solution is stable to all small perturbations keeping the endpoints fixed''. In essence, generalization to the higher dimension does not affect the stability of the gravastar.

\subsection{Size}
To discuss the size of gravastars we follow its phenomenology and start from the interior condensate phase which obey the EOS $p = -\rho$. This is exactly that of the cosmological dark energy which is thought to be responsible for the accelerating phase of the present universe \cite{Riess1998,Perlmutter1999}. It is argued by Mottola \cite{Mottola2001} that the gravastar solution automatically fixes the vacuum condensate energy density ($\rho_{cond}$) in terms of its size. This raised an interesting possibility that the observable universe itself could be the interior of a gravastar. In that case, following  Mottola \cite{Mottola2001} the observed cosmological dark energy of our universe can be calculated as
 \begin{equation}
\rho_{cond} \simeq \ 72\ \%\  of \rho_{crit}=(0.72) \frac{3 c^4 H_0^2}{8\pi G}\simeq 6.9\times 10^{-9}~erg/cm^3,\label{eq51}
\end{equation}
where it is assumed that some $72 \%$ of the critical energy density $\rho_{crit}$ defined by the present value of the Hubble
parameter $H_0$ would be identified with the condensate energy density.

Let us now set the energy scale of the interior de Sitter energy density to {\it TeV}, so that from above Eq. (\ref{eq51}) we have
 \begin{equation}
\rho_{cond} \simeq 4.307 \times 10^{6}~Tev/km^3.
\end{equation}

This is a huge energy which is required to get a gravastar that is comparable to the size of the observable universe. One can even find a link of this energy to the scale of inflation or GUT ($10^{16}~GeV$) as the above energy turns out to be $\rho_{cond} \simeq 4.307 \times 10^{9}~Gev/km^3$.

\section{Discussions and Conclusions}
In the present study of gravastars with higher dimensional manifold and in the presence of
cosmological constant, we have considered several aspects of the system. To examine a new
model of gravastar in comparison with the type proposed by Mazur-Mottola~\cite{Mazur2001,Mazur2004},
we are especially searching for its generalization to: (i) the extended $D$
dimensional spacetime from the $4$ dimensions and (ii) the cosmological constant.
It is worth to mention that using the consideration (i) along with the inclusion of Maxwell's spacetime we have
already obtained an interesting class of solutions~\cite{Ghosh2017}. However, in the present work
we have used the consideration (i) combined with the cosmological constant  to explore possibility of a physically viable astrophysical system
which can be considered as an alternative to $D$-dimensional Schwarzschild–Tangherlini
category black hole~\cite{Tangherlini1963}.

In this section we are discussing some key physical features of the model (observed as well as speculated) as follows:

(i) We have obtained several physical parameters, e.g. metric potentials,
proper length of the shell, energy, entropy etc. and our results match
with the results of Usmani et al.~\cite{Usmani2011} without the cosmological
constant in higher dimensional spacetime. The variations of the parameters as
shown in the plots (Figs. 1-5) indicate in favour of the physical acceptability
and for the existence of gravastars.

(ii) We have calculated different parameters for three specific
regions of the gravastar. All the features of the solutions suggest
that the cosmological constant plays an important role for the
construction of gravastars.

(iii) From Eqs. (10), (12), (14) and Fig. 1  we can observe that the pressure,
 matter density and the metric potentials  are finite at the centre
 (i.e. $r=0$) of the gravastar. So the solutions that we have obtained
 are completely singularity free and also maintain regularity conditions
 inside the star.

(iv) The variation of the energy and the entropy is shown in Fig. 3 and
Fig. 4 respectively. From these figures it can be observed that both energy
and entropy are decreasing with the dimensions. This in turn indicates that
the shell becomes less compact and the matter density must decrease with
dimensions. Exactly the same nature of variation for the matter density
within the shell has been observed in Fig. 2.

(v) From Fig. 5 we note that the proper length is decreasing
with the increase in dimensions, which suggests that the shell become thinner
in higher dimensions.  This is again well justification of the Fig. 2,
i.e. as the matter density decreases the energy as well as the entropy
will also decrease.

(vi) In connection to Figs. 1 and 2 let us now look at the issue: how large the star will be as a function of the energy density and number of dimensions. For the interior region, the equation of state $\rho = \rho_c$ (Eq. (10)) does not contribute anything to the size of the gravastar. However, via Eq. (13) we note from Fig. 1 that the size of the gravastar depends entirely on the increasing order of dimensions. In the case of intermediate region, the thin shell follows the expression (23) which is characterized by Fig. 2 and indicates that the density decreases with increasing mode of dimension. Hence, the total mass of the gravastar being constant the volume of the star becomes larger with decreasing density. Thus, for both the cases we observe that the size of the gravastar increases with higher order of dimensions.

(vii) In comparison with the previous work by Ghosh et al. \cite{Ghosh2017}
it can be seen that the presence of cosmological constant for the construction
of gravastar in higher dimensions shows almost the similar effects as observed in
the presence of charge in the gravastar. This observation therefore indicates that
the repulsive nature of the Colombian charge and cosmological constant have
the similar role as far as gravastar is concerned.

(viii) It is observed that all the obtained results of the present model on gravastar
are overall very much indicative that higher dimensional approach
to construct a gravastar seems theoretically sound and solutions are
physically acceptable. However, a close observation of different profiles and
plots show that the higher dimensional modelling of a gravastar
does not indicate any significant difference in nature of the physical parameters
from that of the ordinary four dimensional spacetime with or without cosmological constant.

As a final comment we would like to mention of a methodology which is instructive to
prove the existence of a gravastar in comparison to black hole. The scheme has been
first developed by Kubo and Sakai~\cite{KS2016} for the case of spherical thin-shell model
of a gravastar provided by Visser and Wiltshire~\cite{VW2004} and later on have been exhibited
by Das et al.~\cite{Das2017}. The method is involved in finding out microlensing effects
for the gravastar, where the maximal luminosity could be considerably larger than
the black hole with the same mass.

In this context, one may raise the following issue: Could gravastar be a substantial fraction of the dark matter while evading microlensing constraints? As gravastar is an alternative to black hole and at present we do not have conclusive evidences of gravastar, so the explanation may be suitable in terms of rather black hole. The primordial black holes are considered to be one of the plausible candidates of dark matter which may be detected through the gravitational micro-lensing effect. The EROS and MACHO surveys have put a limit on the abundance of primordial black holes in the range $10^{23} - 10^{31}$~kg which indicate that primordial black holes within this range cannot constitute an important fraction of the dark matter~\cite{Tisserand2007,Alcock1998}. However, the micro-lensing constraints could naturally be evaded in the case of regrouping of primordial black holes in dense halos~\cite{Clesse2017}. As these surveys and information related to primordial black holes may be put as input in the case study of gravastars.

\section*{Acknowledgements}
The authors FR and SR are grateful to the authority of
Inter-University Center for Astronomy and Astrophysics, Pune, India
for providing Associateship under which a part of this work was carried out.
SR is also grateful to the authority of The Institute of Mathematical Sciences,
Chennai, India for the assistance and facilities. The authors all are grateful to
the anonymous referee for the pertinent comments which
have helped them to upgrade the manuscript substantially.


\begin{thebibliography}{99}

\bibitem{Mazur2001} P. Mazur, E. Mottola, arXiv:gr-qc/0109035, Report number: LA-UR-01-5067 (2001)

\bibitem{Mazur2004} P. Mazur, E. Mottola, Proc. Natl. Acad. Sci. USA \textbf{101}, 9545 (2004)

\bibitem{Visser2004} M. Visser et al., Class. Quantum Gravit. \textbf {21}, 1135 (2004)

\bibitem{Cattoen2005} C. Cattoen, T. Faber, M. Visser, Class. Quantum Gravit. \textbf {22}, 4189 (2005)

\bibitem{Carter2005} B.M.N. Carter, Class. Quantum Gravit. \textbf{22}, 4551 (2005)

\bibitem{Bilic2006} N. Bilic et al., JCAP \textbf{0602}, 013 (2006)

\bibitem{Lobo2006} F. Lobo, Class. Quantum Gravit. \textbf{23}, 1525 (2006)

\bibitem{DeBenedictis2006} A. DeBenedictis et al., Class. Quantum Gravit. \textbf{23}, 2303 (2006)

\bibitem{Lobo2007} F. Lobo et al., Class. Quantum Gravit. \textbf{24}, 1069 (2007)

\bibitem{Horvat2007} D. Horvat et al., Class. Quantum Gravit. \textbf{24}, 4191 (2007)

\bibitem{Cecilia2007} B.M.H. Cecilia et al., Class. Quantum Gravit. \textbf{24}, 5637 (2007)

\bibitem{Rocha2008} P. Rocha et al., J. Cosmol. Astropart. Phys. \textbf{11}, 010 (2008)

\bibitem{Horvat2008} D. Horvat, S. Ilijic and A. Marunovic, Class. Quantum Gravit. \textbf{26}, 025003 (2009)

\bibitem{Nandi2009} K.K. Nandi et al., Phys. Rev. D {\bf 79}, 024011 (2009)

\bibitem{Turimov2009} B.V. Turimov, B.J. Ahmedov, A.A. Abdujabbarov, Mod. Phys. Lett. A \textbf{24}, 733 (2009)

\bibitem{Usmani2011} A.A. Usmani et al., Phys. Lett. B \textbf{701}, 388 (2011)

\bibitem{Rahaman2012a} F. Rahaman et al., Phys. Lett. B \textbf{707}, 319 (2012)

\bibitem{Rahaman2012b} F. Rahaman et al., Int J Theor Phys \textbf{54}, 50 (2015)

\bibitem{Rahaman2012c} F. Rahaman et al., Phys. Lett. B \textbf{717}, 1 (2012)

\bibitem{Ghosh2017} S. Ghosh, F. Rahaman, B.K. Guha, and S. Ray, Phys. Lett. B \textbf{767}, 380 (2017).

\bibitem{Riess1998} A.G. Riess et al., Astron. J. \textbf{116}, 1009 (1998)

\bibitem{Perlmutter1999} S. Perlmutter et al., Astrophys. J. \textbf{517}, 565 (1999)

\bibitem{Ray2007b} S. Ray, U. Mukhopadhyay, X.-H. Meng, Gravit. Cosmol. \textbf{13}, 142 (2007)

\bibitem{Usmani2008} A.A. Usmani, P.P. Ghosh, U. Mukhopadhyay, P.C. Ray, S. Ray, Mon. Not. R. Astron. Soc. \textbf{386}, L92 (2008)

\bibitem{Frieman2008} J. Frieman, M. Turner, D. Huterer, Ann. Rev. Astron. Astrophys. \textbf{46}, 385 (2008)

\bibitem{Davies1984} P.C.W. Davies, Phys. Rev. D \textbf{30}, 737 (1984)

\bibitem{Blome1984} J.J. Blome, W Priester, Naturwis. \textbf{71}, 528 (1984)

\bibitem{Hogan1984} C. Hogan, Nat. \textbf{310}, 365 (1984)

\bibitem{Kaiser1984} N. Kaiser, A. Stebbins, Nat. \textbf{310}, 391 (1984)

\bibitem{Lobo2008} F.S.N. Lobo, arXiv:0805.2309 [gr-qc]

\bibitem{Chan2009a} R. Chan, M.F.A. da Silva, J.F.V. da Rocha, Gen. Relativ. Gravit. \textbf{41}, 1835 (2009)

\bibitem{Chan2009b} R. Chan, M.F.A. da Silva, J.F.V. da Rocha, Mod. Phys. Lett. A \textbf{24}, 1137 (2009)

\bibitem{Zeldovich1972} Y.B. Zel'dovich, Mon. Not. R. Astron. Soc. \textbf{160}, 1 (1972)

\bibitem{Carr1975} B.J. Carr, Astrophys. J. \textbf{201}, 1 (1975)

\bibitem{Madsen1992} M.S. Madsen, J.P. Mimoso, J.A. Butcher, G.F.R. Ellis, Phys. Rev. D \textbf{46}, 1399 (1992)

\bibitem{Buchert2001} T. Buchert, Gen. Relativ. Gravit. \textbf{33}, 1381 (2001)

\bibitem{Braje2002} T.M. Braje, R.W. Romani, Astrophys. J. \textbf{580}, 1043 (2002)

\bibitem{Linares2004} L.P. Linares, M. Malheiro, S. Ray, Int. J. Mod. Phys. D \textbf{13}, 1355 (2004)

\bibitem{Hubble1929} E.P. Hubble,  Proc. Natl. Acad. Sci. \textbf{15}, 168 (1929)

\bibitem{Casimir1948} H.B.G. Casimir, Kon. Ned. Akad. Wetensch. Proc. \textbf{51}, 793 (1948)

\bibitem{Ruderman1972}  R. Ruderman, Rev. Astr. Astrophys. \textbf{10}, 427 (1972)

\bibitem{Perlmutter1998} S. Perlmutter et al., Nat. \textbf{391}, 51 (1998)

\bibitem{Schwarz1985} J.H. Schwarz, Superstings, World Scientific, Singapore (1985)

\bibitem{Weinberg1986} S. Weinberg, Strings and Superstrings, World Scientific, Singapore (1986)

\bibitem{Duff1995} M.J. Duff, J.T. Liu, R. Minasian, Nucl. Phys. B {\bf 452}, 261 (1995)

\bibitem{Polchinski1998} J. Polchinski, String Theory, Cambridge (1998)

\bibitem{Hellerman2007} S. Hellerman, I. Swanson, J. High Energy Phys. {\bf 0709}, 096 (2007)

\bibitem{Aharony2007} O. Aharony,  E. Silverstein, Phys. Rev. D {\bf 75}, 046003 (2007)

\bibitem{Bhar2014} P. Bhar, Astrophys. Space Sci. \textbf{354}, 457 (2014)

\bibitem{Pradhan2007}  A. Pradhan,  G.S. Khadekar,  M.K. Mishra,  S. Kumbhare, Chin. Phys. Letter. {\bf 24}, 3013 (2007)

\bibitem{Emparan2008}  R. Emparan, H.S. Reall, Liv. Rev. Relativ. {\bf 11}, 6 (2008)

\bibitem{Khadekar2014} G.S. Khadekar, S. Kumbhare, Clifford Analysis, Clifford Algebras and Their Applications {\bf 3}, 307 (2014)

\bibitem{Tangherlini1963} F.R. Tangherlini, Nuo. Cim. \textbf{27}, 636 (1963)

\bibitem{Darmois1927} G. Darmois, “M{\'e}morial des sciences math{\'e}matiques XXV”,
Fasticule XXV, chap. V (Gauthier-Villars, Paris, France, 1927)

\bibitem{Israel1966} W. Israel, Nuo. Cim. \textbf{66}, 1 (1966)

\bibitem{Lanczos1924} C. Lanczos, Ann. Phys. (Leipzig) \textbf{74}, 518 (1924)

\bibitem{Mottola2001} E. Mottola, J. Phys.: Conf. Ser. \textbf{314}, 012010, (2011)

\bibitem{KS2016} T. Kubo, N. Sakai, Phys. Rev. D \textbf{93}, 084051 (2016)

\bibitem{VW2004} M. Visser, D.L. Wiltshire, Class. Quantum Gravit. \textbf{21}, 1135 (2004)

\bibitem{Das2017} A. Das, S. Ghosh, B.K. Guha, S. Das, F. Rahaman, S. Ray, Phys. Rev. D \textbf{95}, 124011 (2017)

\bibitem{Tisserand2007} P. Tisserand, et al., Astron. Astrophys. \textbf{469}, 387 (2007)

\bibitem{Alcock1998} C. Alcock, et al., Astrophys. J. \textbf{499}, L9 (1998)

\bibitem{Clesse2017} S. Clesse, J. Garcia-Bellido, Phys. Dark Univ. \textbf{15}, 142 (2017)

\end{thebibliography}
\end{document}